\documentclass[12pt]{article}
\usepackage{amsfonts,amsmath,epsfig,graphicx,amssymb}
\usepackage{amsmath,epsfig,graphicx,amssymb}
\usepackage[margin=1in]{geometry}
\usepackage{makeidx}
\usepackage{float}
\usepackage{xurl}
\usepackage{hyperref}
\usepackage{booktabs}
\usepackage{hyperref}
\usepackage{xurl}
\newcommand{\beq}{\begin{equation}}
\newcommand{\eeq}{\end{equation}}
\newcommand{\ben}{\begin{eqnarray}}
\newcommand{\een}{\end{eqnarray}}
\date{}
\begin{document}
\title{Testing quantum-like markers in neural dynamics}
\author{Partha Ghose$^1$\footnote{\url{partha.ghose@gmail.com}}\\
$^1$Tagore Centre for Natural Sciences and Philosophy,\\
Rabindra Tirtha, New Town, Kolkata 700156, India\\
Dimitris Pinotsis$^2$\footnote{\url{pinotsis@mit.edu}}\\
$^2$Centre for Clinical, Social and Cognitive Neuroscience and\\
Department of Psychology and Neuroscience,\\
City St George's, University of London,\\ London EC1V 0HB, United Kingdom}   

\maketitle
\begin{abstract}
We propose two experiments for identifying quantum markers in neural data based on quantum variants of well-known equations for neural activity that describe electrical signal propagation on axonal arbors and dendrites. These include (i) testing if power spectra from subthreshold oscillations in neuronal cultures follow the classical Fitzgugh-Nagumo equations or a recently introduced quantum variant of them and (ii) testing if propagation statistics of electrical activity in axons follow the classical diffusive cable equation or a quantum variant of it.

\end{abstract}

\section{Introduction}

The boundary between classical and quantum descriptions of natural phenomena has long been a subject of foundational interest. While quantum mechanics is traditionally viewed as a fundamental theory for microscopic systems, several lines of recent research have explored whether quantum-like behaviour might emerge from classical stochastic dynamics \cite{Jedlicka,AdamsPetruccione}. 

For decades, most neuroscientists and physicists dismissed the idea that quantum processes could influence brain function, arguing that the warm, wet, and noisy conditions of macroscopic biological systems would suppress fragile quantum states. But recent advances in quantum biology have prompted a reevaluation of that assumption \cite{lamb, mac}.

In the realm of neuroscience, growing interest surrounds the possibility that certain mental states underlying cognition exhibit quantum effects. These include non-commutative probabilities, contextuality, and interference effects \cite{bus, khren, pothos, aerts, true, conte, asano, bruza}. In a different line of work \cite{ham}, quantum effects in the cytoskeleton and the microtubules have been proposed as a substrate for consciousness.  However, there is no evidence for fundamental quantum effects at the neural level. 

In our earlier paper \textit{``The FitzHugh-Nagumo equations and quantum noise''} \cite{ghose}, it was shown that the dynamics of excitable neurons described by the FitzHugh-Nagumo (FN) model \cite{fitzhugh1, nagumo} with Gaussian noise \emph{can} indeed be recast into a quantum Schr\"{o}dinger form, using the methods introduced by Nelson and others \cite{Nelson,GuerraMorato}. A new  neuronal constant, functionally analogous to Planck’s constant, hinting at the emergence of a quantized model for neural processes, was also proposed. This established a \emph{mathematical} link between neuronal models and quantum  equations. Given that neuronal models describe brain activity underlying cognition, this link also suggests that quantum effects \emph{could} underlie cognition. In this setting quantum-like coherence `emerges' from electric fields at the mesoscale, i.e. in data recordings obtained with electrophysiology and similar brain imaging techniques.

A crucial ingredient in recasting the classical FN equations into a Schr\"{o}dinger equation was the incorporation of Gaussian (Wiener) noise \cite{Nelson,GuerraMorato}. This noise structure was then shown to be related to quantum effects described by a wave function \cite{Nelson2,Susskind,Ghirardi}. \emph{Here we show that another noise process known as Kac-type velocity-jump process with finite propagation speed is related to another quantum wave equation, the Dirac equation}. Excitatory postsynaptic potentials (EPSPs) and inhibitory postsynaptic potentials (IPSPs) are known to generate excursions of the membrane potential respectively toward and away from the firing threshold. \emph{We describe these processes by the forward and backward sectors of a Kac-type persistent stochastic process}. Based on that, we propose two critical experiments to distinguish between classical and such quantum-like models of neural processes. In one experiment, we propose a test for quantum–like dynamics in spectral responses emitted from subthreshold oscillations. In a second experiment we propose a test to distinguish between Kac-type persistent processes and diffusive propagation.

\subsection*{Scope of the term `quantum'}

In the present work, the term ``quantum'' is used in a strictly operational and effective sense. 
While classical stochastic dynamics can, under suitable transformations, be recast into Schr\"odinger- or Dirac-like forms, the experimental programme proposed here does not target interference or coherence phenomena, as such signatures can admit classical interpretations and are therefore not diagnostically decisive.
Instead, it focuses on two measurable signatures as `quantum markers': 
(i) the existence of an effective neural constant $\hat{\hbar}$ inferred from fluctuation statistics around subthreshold activity, and 
(ii) the scaling behaviour of propagation statistics (arrival-time distributions and timing dispersion) that distinguishes diffusive from finite-speed persistent transport. 

In this sense, ``quantum'' refers to the structure of the effective description and the associated measurable parameters, rather than to a claim about the microscopic physical nature of neural processes.

We will first review the basics of the classical model in section 2, and then introduce the Kac process from statistical physics in section 3. The experimental proposals will be explained and discussed in section 4.

\section{Classical Model}
Let us briefly summarize the main features of the classical model of signal propagation in neurons.
The analogy between telegraph wires and neurons dates back to the work of Lord Kelvin and later Hermann Helmholtz, but it was Hodgkin and Huxley \cite{hh} who formalized it in the early 1950s, leading to their Nobel Prize. They modeled the axon as a distributed electrical circuit. This was essentially a biological instantiation of the Telegrapher's equation.

To avoid any terminological confusion, let us start by defining what we will mean by the ``Telegrapher's equation'' and the ``cable equation'' in  this paper. The Telegrapher's equation (in transmission line theory) is the more general line model that includes inductance (RLGC line). It is a hyperbolic/wave-like PDE (second order in both space and time) with finite propagation speed. In computational neuroscience, the (passive) cable equation is the standard parabolic PDE (first-order in time, second-order in space) for voltage spread along dendrites and axons modeled as RC distributed lines (axial resistance and membrane capacitance/leak).   

In modern computational neuroscience, stochastic versions of the Telegrapher's equation, often under different names, are used to model ion channel noise, spike propagation variability, and stochastic resonance \cite{faisal, tuck, gold, gam}. Some models attempt to go beyond the purely diffusive cable model to include wave-like effects, especially in the context of demyelinated axons or electromagnetic pulse propagation.

In its simplest passive form, the cable equation is a diffusion-like PDE for the membrane potential
$V(x,t)$ along a one-dimensional cable:
\begin{equation}
C_m \,\partial_t V
= D\,\partial_x^2 V - g_L\,(V-E_L),
\label{eq:cable}
\end{equation}
where $C_m$ is membrane capacitance per unit length, $D$ is the effective axial-diffusion coefficient
(set by axial resistance and geometry), $g_L$ is the leak conductance, $E_L$ denotes the leak reversal (resting) potential, i.e.\ the voltage at which the passive leak current vanishes. 

If one models spatial voltage propagation along an axon or dendrite by a finite-speed transport law with effective propagation speed $v$, then a classical Telegrapher-type equation for the membrane potential $V(x,t)$ may be written as
\begin{equation} 
\partial_t^2 V + 2\gamma\,\partial_t V = v^2\,\partial_x^2 V, \label{eq:telegraph} 
\end{equation} 
where $\gamma$ is the damping rate and $E_L$ is the resting (leak) potential. 
In the long-time regime, $t\gg \gamma^{-1}$, the inertial term $\partial_t^2 V$ becomes negligible compared with $2\gamma\,\partial_t V$ (see below), and (\ref{eq:telegraph}) reduces to the diffusive cable-type form
\begin{equation} 
\partial_t V \approx \frac{v^2}{2\gamma}\,\partial_x^2 V.\label{diffcable}
\end{equation} 
The neglect of the term $\partial_t^2 V$ at long times follows from a standard separation of
time scales. If the voltage varies over a characteristic time scale $T$, then
\[
\partial_t V \sim \frac{V}{T},
\qquad
\partial_t^2 V \sim \frac{V}{T^2}.
\]
Hence the ratio of the inertial term to the damping term is
\[
\frac{|\partial_t^2 V|}{|2\gamma\,\partial_t V|}
\sim
\frac{1}{2\gamma T}.
\]
Therefore, in the regime $T \gg \gamma^{-1}$, or equivalently for times much longer than the
damping time $\gamma^{-1}$, one has
\[
|\partial_t^2 V| \ll |2\gamma\,\partial_t V|,
\]
so the second time derivative becomes negligible and Eq. (\ref{eq:telegraph}) reduces to the diffusive form
given in Eq. (\ref{diffcable}).

Thus the Telegrapher's equation provides a finite-speed extension of the classical cable equation, with the cable limit recovered for $t \gg\gamma^{-1}$ in saltatory conduction with $D/C_m = v^2/2\gamma$.

\section{ Kac-type persistent random walk and telegrapher coupling}
A Kac-type persistent random walk \cite{kac} models transport in \emph{any} system at finite speed $v$ with direction
reversals at a Poisson rate $\lambda$. (The Poisson rate $\lambda$ is the average switching probability per unit time:
during a short interval $dt$, the probability of a reversal is $\lambda\,dt$.
Equivalently, $\lambda^{-1}$ is the mean time for which propagation persists in
one direction before being randomized by a reversal.)   In a 1D setting one writes a two-component transport system for
right/left movers (purely as direction labels):
\begin{equation}
\partial_t P_\pm(x,t)
= \mp v\,\partial_x P_\pm(x,t) \;-\; \lambda\bigl(P_\pm(x,t)-P_\mp(x,t)\bigr).
\label{eq:kac}
\end{equation}
where $P_{\pm}(x,t)$ are probability densities describing the dependence of the membrane potential over space as it propagates in axons and dendrites . Starting from this equation, define the total density and current
\[
P=P_+ + P_-,
\qquad
J=P_+ - P_-.
\]
Then adding and subtracting the two equations gives
\[
\partial_t P = -\,v\,\partial_x J,
\qquad
\partial_t J = -\,v\,\partial_x P - 2\lambda J.
\]
Differentiating the first equation with respect to time,
\[
\partial_t^2 P = -\,v\,\partial_x(\partial_t J).
\]
Using the second equation,
\[
\partial_t^2 P
=
-\,v\,\partial_x\bigl(-\,v\,\partial_x P - 2\lambda J\bigr)
=
v^2\,\partial_x^2 P + 2\lambda v\,\partial_x J.
\]
Finally, since $\partial_t P = -\,v\,\partial_x J$, we obtain
\[
\partial_t^2 P
=
v^2\,\partial_x^2 P - 2\lambda\,\partial_t P,
\]
or equivalently,
\begin{equation}
(\partial_t^2  + 2\lambda\,\partial_t - v^2\,\partial_x^2) P = 0. \label{eq:telegrapher}
\end{equation}
Thus, the total probability density $P=P_+ + P_-$ satisfies the Telegrapher's equation.

Although the Telegrapher equations for the membrane potential $V(x,t)$
and the probability density $P(x,t)$ have the same mathematical form,
their physical interpretations are fundamentally different. In the
former case, the equation is a phenomenological description of finite-speed
signal propagation in neural tissue, with $V$ representing a directly
measurable membrane potential. In the latter case, the equation emerges
from an underlying Kac-type persistent stochastic process, where
$P$ denotes a probability density and the parameters $v$ and
$\lambda$ have the explicit interpretation of propagation speed and
stochastic reversal rate. Thus identical mathematical equations arise
from distinct physical assumptions: a continuum transport description in
one case and a persistent random-walk dynamics in the other.

The only important changes relative to \eqref{eq:cable} are{} the presence of a second-order
time derivative and a finite-speed transport scale $v$.

Since $P$ is the probability density for the position of a persistent random walker,
the operator appearing in equation (\ref{eq:telegrapher}) also acts as the \emph{propagator}
(Green’s function) for any quantity transported by the same finite-speed stochastic mechanism.
In the regime
\begin{equation}
t \gg \lambda^{-1},
\label{eq:diff_limit_cond}
\end{equation}
the term $\partial_t^2 P$ in \eqref{eq:telegrapher} becomes negligible compared to
$2\lambda\,\partial_t P$, giving
\begin{equation}
\partial_t P \approx \frac{v^2}{2\lambda}\,\partial_x^2 P,
\qquad
D_{\mathrm{eff}}=\frac{v^2}{2\lambda}.
\label{eq:diff_limit}
\end{equation}
Thus the telegrapher model \emph{contains} the diffusion/cable behaviour in the coarse-grained (diffusive) limit, valid when one probes only time scales $t \gg \lambda^{-1}$, so that the short-time directional reversals are averaged out. In this limit $\lambda$ is proportional to the classical damping rate $\gamma$ in eqn (\ref{diffcable}).
This is crucial for the experimental tests: it means the classical cable equation is recovered at long scales,
while the stochastic model predicts additional structure at short scales (where the approximation
\eqref{eq:diff_limit_cond} fails and induction effects are important).

Note that equation \eqref{eq:kac} is a real, positivity-preserving Markov evolution for probabilities and
therefore does not by itself support interference-like phase phenomena.  Now, as shown by
Gaveau--Jacobson--Kac--Schulman \cite{gav}, after a simple phase transformation
\[
u_{\pm}(x,t) = \exp ({imc^2t/\hbar})\psi_{\pm}(x,t),
\]
the Dirac 
equation can be written in the two-component form 
\begin{equation}
\partial_t u_{\pm} = \frac{imc^2}{\hbar}\left(u_{\pm} - u_{\mp}\right) \mp c\partial_x u_{\pm} 
\end{equation}
which matches equation \eqref{eq:kac} under the following identifications: 
\begin{equation}
 c \leftrightarrow v,
\qquad
\lambda \leftrightarrow -\,\frac{i m c^2}{\hbar},
\quad 
u_{\pm} \leftrightarrow   P_{\pm}.
\label{eq:analytic_cont} 
\end{equation}
This identification is not a mere relabelling: it converts a real relaxation generator into a purely
imaginary phase generator, thereby introducing a genuine complex amplitude. The connection between the two descriptions is thus mathematical and formal, not physical.

The Schr\"{o}dinger dynamics can be viewed either as an envelope/nonrelativistic limit of Dirac-type dynamics or as the natural complex-amplitude reorganization of an underlying Wiener (diffusion) process, as in our earlier approach \cite{ghose}. There, starting from a random walk model with drift \cite{GM1964}, we applied stochastic mechanics—specifically the
method developed by Nelson—to reinterpret the neural dynamics in a form analogous to
quantum mechanics. Within this framework, the probabilistic evolution of the neuron’s
state was reformulated as a Schr\"{o}dinger-like equation defined over the wave function
$\psi(q,t)$ where $q$ was the state variable and $|\psi(q,t)|^2$ was the probability density.
In other words, the dynamics of the probability distribution governing the neuron's
behaviour took on a mathematical structure similar to the Schr\"{o}dinger equation, even
though the system itself could remain classical and driven by noise rather than quantum
effects. Besides random walk models, in \cite{ghose} we also showed that this correspondence was
also true for the well known FitzHugh–Nagumo model with stochastic noise.

While our earlier research paper focused on the emergence of a Schr\"{o}dinger-type description from Gaussian (Wiener) noise and introduced an effective neural Planck constant $\hat{\hbar}$ in connection with stochastic FitzHugh–Nagumo dynamics, \emph{here we introduce an alternative and fundamentally different stochastic mechanism, namely a Kac-type velocity--jump process with finite propagation speed. This possibility is motivated by the physiological observation that excitatory and inhibitory postsynaptic potentials (EPSPs and IPSPs) naturally generate excursions of the membrane potential respectively toward and away from firing threshold, suggesting an underlying forward--backward stochastic dynamics}. The resulting continuum description is governed by the Telegrapher equation and is formally related to the Dirac equation through the construction of Gaveau, Jacobson, Kac, and Schulman \cite{gav}. \emph{Unlike the earlier Wiener-based framework, the Kac formulation predicts finite-speed transport, persistence effects, and experimentally testable arrival-time and first-passage statistics}, as we will show in the next section.
\section{Proposed Experiments}

We propose two types of experiment: (a) measurement of the effective neural Planck constant $\hat{\hbar}$ and (b) distinguishing between passive cable and Kac-type random-walk transport models.

\subsection*{Experiment I: subthreshold oscillations and effective $\hat{\hbar}$ scaling}

\subsubsection*{Experimental idea and implementation}
We will measure a proxy of the thermal energy emitted in a neuronal culture as a result of intracellular membrane potential subthreshold oscillations. We will assess if thermal energy is better predicted by a classical or a quantum variant of the Fitzgugh-Nagumo model introduced in \cite{ghose}. Note that, following the link between the Fitzgugh-Nagumo and Schr\"{o}dinger equation introduced in  [12], this experiment will focus on the  statistics of fluctuations around sub-threshold oscillations, not the  subthreshold oscillations themselves .

The proposed protocol is:

\begin{enumerate}
    \item \emph{Perform patch-clamp recordings} of membrane potential in neuronal cultures exhibiting subthreshold oscillations (for example, magnocellular neurons in the spirit of \cite{schmitz}, or a comparable preparation), using established whole-cell methods \cite{neher}.
    \item \emph{Monitor temperature} in the same culture, using thermocouples \cite{lee}. Note that temperature sets the thermal baseline against which any effective-$\hat{\hbar}$ scaling must be tested.
    \item \emph{Record spectral power} of neural activity following standard methods \cite{welch1967, pinotsis1,pinotsis2}.
\end{enumerate}

Note that subthreshold oscillations are near-threshold rhythmic variations of membrane depolarization potential that can be described by classical models like Fitzgugh-Nagumo and similar models (a dynamical feature of the deterministic or weakly noisy neuronal system) \cite{schmitz}. On top of these deterministic variations there can be some stochastic noise, as a result of random synaptic input, thermal effects, instrumentation limitations, etc. \cite{rudolph,faisal,gold} that we call fluctuations or noise (see figure below). 

\begin{figure}[H]
\centering
\includegraphics[width=0.9\linewidth]{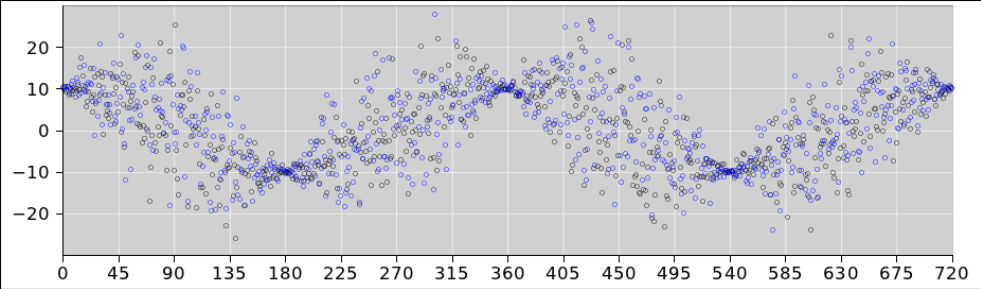}
\caption{Schematic illustration of experimentally observed subthreshold
oscillations and stochastic fluctuations of the membrane potential
(reproduced from Ref \cite{ghose}). The figure is repeated here for the reader's convenience. In Ref. \cite{ghose} these fluctuations were described within a Wiener-noise framework, but no attempt was made to estimate the value of the effective Planck constant $\hat{\hbar}$. In contrast, the present work proposes a method for estimating its value.}
\label{fig:scatter}
\end{figure}

\subsubsection*{How $\hat{\hbar}$ is extracted}

Like we said in \cite{ghose}, subthreshold oscillations were described by a classical Fitzgugh-Nagumo model of neural activity (with additive stochastic noise) and a quantum Schr\"{o}dinger equation that was derived from it. This is a quantum variant of the classical FN model. It was also shown that the effective Planck constant $\hat{\hbar}$ appearing in the new Schr\"{o}dinger equation can be obtained from the variance of noise (fluctuations) around the   
oscillating membrane potential.

\subsubsection*{Energy--frequency relation and what is actually tested}

Given the effective Planck constant, \cite{ghose} obtained the thermally averaged energy corresponding to subthreshold oscillations of membrane depolarization in neuronal cultures given by the equation:
\begin{equation}
\langle E\rangle=\frac{\hat{\hbar}\omega}{2}+\frac{\hat{\hbar}\omega}{e^{\beta \hat{\hbar}\omega}-1},
\qquad \beta=\frac{1}{k_B T}.
\label{aven}
\end{equation}

In the experiment we suggest here one does \emph{not} directly measure $\langle E\rangle$ at the microscopic level. Instead, one uses an experimentally accessible proxy derived from the membrane-potential time series, such as integrated spectral power in a narrow band around the subthreshold peak frequency,
\begin{equation}
P_{\mathcal{B}}=\sum_{f\in\mathcal{B}} S_{VV}(f)\,\Delta f,
\end{equation}
with $\mathcal{B}$ centered on the subthreshold oscillation peak. The testable claim is to compare the frequency- and temperature-dependence of $P_{\mathcal{B}}$ against the energy described by equation (\ref{aven}) that follows from the quantum variant of the FN model.

Alternatively, the well-known classical FN model predicts that the spectral density of subthreshold oscillations is given by the classical expression 

\begin{equation}
S_x(\omega) = (i\omega \mathbb{I} - J)^{-1} BQB^T (-i\omega \mathbb{I} - J^T)^{-1}
\label{FN}
\end{equation}
where $J$ is the Jacobian matrix, $B$ is the noise coupling matrix and $Q$ is the noise covariance. Then, data recorded in neuronal cultures and model fitting will reveal if the classical or quantum expressions given respectively by equations (\ref{FN}) or (\ref{aven}) provides a better data fit.

\subsection*{Experiment II:  Passive Cable Propagation vs.\ Kac-Type Random-walk Transport}
The central idea can be stated as follows: the classical cable equation \eqref{eq:cable} is the established diffusion baseline for spatial voltage spread.  A Kac-type persistent random walk yields a
finite-speed alternative \eqref{eq:telegraph} whose diffusive limit
\eqref{eq:diff_limit} recovers the cable equation with $\lambda \propto \gamma$.  Therefore
the two models can be matched at long scales but differ at short times/short distances.  A
comparative study of observables sensitive to this regime---particularly impulse responses and
arrival-time distributions---can discriminate
which model is physically more accurate in the regime of interest.

In our second experiment we will test whether propagation statistics in controlled axonal geometries are better explained by (i) a passive, diffusion-like cable model given by eqn (\ref{eq:cable}) or (ii) a Kac-type persistent (velocity-jump) model
whose continuum description is the quantum cable equation (\ref{eq:diff_limit}). In the former case one expects signal spread broadening with distance as a result of pure diffusion and the absence of a ballistic front. In the latter case, while the observed coarse-grained behaviour may appear to be diffusive, the underlying process does not have to be standard Brownian diffusion—a persistent (finite-speed) random walk can produce different transient signatures revealing quantum markers. These are explained below.

\subsection*{Experimental idea and inplementation}

Use a microfabricated straight channel that guides a \emph{single axon} (or effectively 1D bundle)
from an input region to multiple recording sites at distances
\[
0<L_1<L_2<L_3 \quad (\text{e.g., } L_i \sim 100\text{--}1000~\mu\mathrm{m} \text{ depending on preparation}).
\]
Record at each $L_i$ using MEA electrodes.

Deliver isolated events at low rate so trials are independent:

\noindent
(i) either single action potentials (brief somatic current pulse or optogenetic pulse), or

\noindent
(ii) standardized subthreshold waveforms (brief pulse/chirp) for frequency-domain system identification.

Time-stamp each input event ($t_0$) using the stimulus trigger and/or the proximal electrode.

\subsection*{Primary observables and predicted qualitative differences}

\subsection*{Observable 1: Arrival-time distributions $p(t;L)$}
For each distance $L_i$, detect the event arrival time $\hat{t}(L_i)$ relative to $t_0$ (e.g.\ first threshold crossing,
peak time in a window, or matched-filter detection), and estimate the histogram/empirical density
\[
p(t;L_i)\quad \text{from many trials.}
\]
A Kac/persistent model predicts an appreciable contribution near the ``ballistic'' time
\[
t \approx \frac{L}{v}\quad (\text{few/no reversals}),
\]
along with a longer tail from multiple reversals.
A diffusion/cable baseline produces broader, purely dispersive spread without a sharp ballistic component.

\subsection*{Observable 2: Distance scaling of timing dispersion}
Compute, for each $L_i$, summary statistics such as mean and variance:
\[
\mu_t(L_i)=\mathbb{E}[\hat{t}(L_i)],\qquad \sigma_t^2(L_i)=\mathrm{Var}(\hat{t}(L_i)).
\]
\textbf{Key discriminator (qualitative):}
For persistent random-walk, propagation has a characteristic finite-speed scale and tends to show
approximately linear-in-$L$ timing structure over an intermediate window,
while in diffusion-like classical propagation timing dispersion tends to grow more strongly with $L$
(e.g.\ broader delays consistent with diffusive scaling).
This is strongest when measured at short distances/times (before any long-time diffusive crossover of the Kac model).

One should fit the Kac/persistent model using one global parameter pair $(v,\lambda)$ to \emph{all} datasets
$\{p(t;L_i)\}_{i=1}^3$ using maximum likelihood or least squares on summary features.
The Kac/persistent model is empirically supported only if:
\begin{enumerate}
\item A single $(v,\lambda)$ provides a good fit \emph{simultaneously} across multiple distances $L_i$
(and ideally across modest changes of stimulus waveform), cf input (ii) above.
\item The fitted parameters are stable under reasonable perturbations (e.g.\ slight changes in detection threshold,
preprocessing band, or trial selection).
\item The passive cable model (with reasonable noise and without ad hoc parameter drift) fails to account for the same multi-$L$ dataset.
\end{enumerate}
If the best-fit $(v,\lambda)$ (or implied $D_{\rm eff}$) is found to drift with $L$ or stimulus condition to preserve fit quality,
this counts against a genuine Kac/persistent mechanism (and suggests the model is being used merely as a flexible curve-fit).

\subsection*{Summary}
In earlier work \cite{ghose} we focused  on stochastic FitzHugh–Nagumo dynamics and an emergent Schr\"{o}dinger description. Here, we  developed a distinct persistent-transport framework leading to new experimental predictions and new tests of quantum-like neural dynamics. This new framework reduces to the former in the diffusive limit and includes two experiments

\begin{enumerate}
 \item \emph{ Experiment I}: Do power spectra produced by subthreshold oscillations in neuronal cultures follow the dynamics of a quantum Schr\"{o}dinger equation derived from a FitzHugh-Nagumo model or is the classical model adequate?
 \item \emph{Experiment II}: Do statistics of electrical signals in neuronal cultures follow a stochastic cable equation \eqref{eq:telegraph} indicating quantum behaviour or the classical passive cable equation \eqref{eq:cable}?   
\end{enumerate}

Even a null result would be scientifically useful, because it would sharpen the domain of validity of standard classical stochastic models in neuronal dynamics. A positive result would motivate a more systematic investigation of the bridge between stochastic physics, effective mesoscopic parameters, and quantum-inspired modelling in neuroscience.
Only if the Kac-type model is vindicated should one proceed to check specific quantum-like `witnesses' like contextuality/Leggett-Garg Inequality violation etc. We intend to develop such proposals in future. 

\subsection*{Acknowledgements}
We thank anonymous reviewers for their critical and helpful comments. We also acknowledge use of AI tools for language polishing.


\begin{thebibliography}{99}
\bibitem{Jedlicka} 
P. Jedlicka, ``Revisiting the Quantum Brain Hypothesis: Toward Quantum (Neuro)biology?'', \emph{Front. Mol. Neurosci.} \textbf{10}, 366 (2017).

\bibitem{AdamsPetruccione}
B. Adams and F. Petruccione, ``Quantum effects in the brain: A review,'' \emph{AVS Quantum Sci.} \textbf{2}, 022901 (2020). 

\bibitem{lamb}
N. Lambert, Y.-N. Chen, Y.-C. Cheng, C.-M. Li, G.-Y. Chen and F. Nori,
``Quantum biology,''
{\em Nature Physics} {\bf 9}, 10--18 (2013).

\bibitem{mac}
J. Al-Khalili and J. McFadden,
{\em Life on the Edge: The Coming of Age of Quantum Biology},
Crown Publishers (2014).

\bibitem{bus}
J. R. Busemeyer and P. D. Bruza,
{\em Quantum Models of Cognition and Decision},
Cambridge University Press (2012).

\bibitem{khren}
A. Y. Khrennikov,
{\em Ubiquitous Quantum Structure: From Psychology to Finance},
Springer (2010).

\bibitem{pothos}
E. M. Pothos and J. R. Busemeyer,
``A quantum probability explanation for violations of `rational' decision theory,''
{\em Proc. R. Soc. B} {\bf 276}(1665), 2171--2178 (2009).

\bibitem{aerts}
D. Aerts and S. Sozzo,
``Quantum Structure in Cognition: Origins, Developments, Successes, and Expectations,''
in {\em The Palgrave Handbook of Quantum Models in Social Science: Applications and Grand Challenges},
E. Haven and A. Khrennikov (eds), Springer (2017).

\bibitem{true}
J. S. Trueblood and J. R. Busemeyer,
``A quantum probability account of order effects in inference,''
{\em Cognitive Science} {\bf 35}(8), 1518--1552 (2011).

\bibitem{conte}
E. Conte, A. Khrennikov, O. Todarello, A. Federici, L. Mendolicchio and J. P. Zbilut, 
``Mental states follow quantum mechanics during perception and cognition of ambiguous figures,'' 
{\em Open Systems \& Information Dynamics} {\bf 16}(1), 85--100 (2009).

\bibitem{asano}
M. Asano, A. Khrennikov, M. Ohya, Y. Tanaka, and I. Yamamoto,
{\em Quantum Adaptivity in Biology: From Genetics to Cognition},
Springer (2015).

\bibitem{bruza}
P. D. Bruza, Z. Wang and J. R. Busemeyer,
``Quantum cognition: a new theoretical approach to psychology,''
{\em Trends in Cognitive Sciences} {\bf 19}(7), 383--393 (2015).

\bibitem{ham}
S. Hameroff and R. Penrose,
``Orchestrated reduction of quantum coherence in brain microtubules: A model for consciousness,''
{\em Mathematics and Computers in Simulation} {\bf 40}(3--4), 453--480 (1996).

\bibitem{ghose}
P. Ghose and D. A. Pinotsis,
``The FitzHugh--Nagumo equations and quantum noise,''
{\em Computational and Structural Biotechnology Journal} {\bf 30}, 12--20 (2025).

\bibitem{fitzhugh1}
R. FitzHugh,
``Mathematical models of excitation and propagation in nerve,''
in {\em Biological Engineering}, H. P. Schwan (ed.),
McGraw--Hill, New York, pp.~1--85 (1969).

\bibitem{nagumo}
J. Nagumo, S. Arimoto and S. Yoshizawa,
``An Active Pulse Transmission Line Simulating Nerve Axon,''
{\em Proc. IRE} {\bf 50}, 2061--2070 (1962).

\bibitem{Nelson}
E. Nelson, ``Derivation of the Schr\"{o}dinger Equation from Newtonian Mechanics,'' \emph{Phys. Rev. Lett.} \textbf{150}, 1079--1085  (1966).

\bibitem{GuerraMorato}
F. Guerra and L. M. Morato, ``Quantization of dynamical systems and stochastic control theory,''  \emph{Phys. Rev. D} \textbf{27}, 1774--1786 (1983).

\bibitem{Nelson2}
E. Nelson, \emph{Quantum Fluctuations}, Princeton University Press (1985).

\bibitem{Susskind}
L. Susskind and A. Friedman, \emph{Quantum Mechanics: The Theoretical Minimum}, Penguin Books (2014).

\bibitem{Ghirardi}
G. Ghirardi, \emph{Sneaking a Look at God's Cards: Unraveling the Mysteries of Quantum Mechanics}, translated by G. Malsbary, Princeton University Press (2021).

\bibitem{hh}
A. L. Hodgkin and A. F. Huxley,
``A quantitative description of membrane current and its application to conduction and excitation in nerve,''
{\em J. Physiol.} {\bf 117}(4), 500--544 (1952).

\bibitem{faisal}
A. A. Faisal, L. P. J. Selen and D. M. Wolpert,
``Noise in the nervous system,''
{\em Nature Reviews Neuroscience} {\bf 9}(4), 292--303 (2008).

\bibitem{tuck}
H. C. Tuckwell,
{\em Introduction to Theoretical Neurobiology, Vol.~1: Linear Cable Theory and Dendritic Structure},
Cambridge University Press (1988).

\bibitem{gold}
J. H. Goldwyn and E. Shea-Brown,
``The What and Where of Adding Channel Noise to the Hodgkin-Huxley Equations,''
{\em PLoS Computational Biology} {\bf 7}(11), e1002247 (2011).

\bibitem{gam}
L. Gammaitoni, P. H\"anggi, P. Jung and F. Marchesoni,
``Stochastic resonance,''
{\em Rev. Mod. Phys.} {\bf 70}(1), 223--287 (1998).


\bibitem{kac}
M. Kac,
``A Stochastic Model Related to the Telegrapher's Equation,''
{\em Rocky Mountain Journal of Mathematics} {\bf 4}(3), 497--509 (1974).

\bibitem{gav}
B. Gaveau, T. Jacobson, M. Kac and L. S. Schulman,
``Relativistic Extension of the Analogy between Quantum Mechanics and Brownian Motion,''
{\em Phys. Rev. Lett.} {\bf 53}(5), 419--422 (1984).

\bibitem{schmitz}
D. Schmitz, T. Gloveli, J. Behr, T. Dugladze and U. Heinemann,
``Subthreshold membrane potential oscillations in neurons of deep layers of the entorhinal cortex,''
{\em Neuroscience} {\bf 85}(4), 999--1004 (1998).

\bibitem{neher}
E. Neher and B. Sakmann,
``The patch clamp technique,''
{\em Scientific American} {\bf 266}(3), 44--51 (1992).

\bibitem{lee}
B. C. Lee, Y. -G. Lim, K. -H. Kim, S. Lee and S. Moon,
``Microfabricated Neural Thermocouple Arrays Probe for Brain Research,''
in {\em TRANSDUCERS 2009 International Solid-State Sensors, Actuators and Microsystems Conference},
Denver, Colorado, IEEE, pp.~338--341 (2009).

\bibitem{welch1967}
P. D. Welch, ``The use of fast Fourier transform for the estimation of power spectra: A method based on time averaging over short, modified periodograms,''
{\em IEEE Transactions on Audio and Electroacoustics} {\bf 15}(2), 70--73 (1967).

\bibitem{pinotsis1}
D. A. Pinotsis,
``Statistical decision theory and multiscale analyses of human brain data,''
{\em Journal of Neuroscience Methods} {\bf 346}, 108912 (2020).

\bibitem{pinotsis2}
D. A. Pinotsis, R. Loonis, A. M. Bastos, E. K. Miller and K. J. Friston,
``Bayesian Modelling of Induced Responses and Neuronal Rhythms,''
{\em Brain Topography} \textbf{32}(4), 569--582 (2019).

\bibitem{rudolph}
M. Rudolph and A. Destexhe,
``A multichannel shot noise approach to describe synaptic background activity in neurons,''
{\em Eur. Phys. J. B} {\bf 52}, 125--132 (2006).

\bibitem{GM1964}
G. L. Gerstein and B. Mandelbrot,``Random Walk Models for the Spike Activity of a Single Neuron'', \emph{Biophysical Journal} \textbf{4}, 41-68 (1964). 

\end{thebibliography}
\end{document}